\documentclass[aps, pra, superscriptaddress, twocolumn,10pt]{revtex4-1}
\usepackage[left=1in, right=1in, top=1in, bottom=1in]{geometry}

\usepackage{bm}
\usepackage{tikz,pgfplots}
\usepackage{standalone}
\usepackage{braket}
\usepackage[retainorgcmds]{IEEEtrantools}
\usepackage{graphicx}
\usepackage{mathrsfs}
\usepackage{amsmath}
\usepackage{amssymb}
\usepackage{color}
\usepackage{amsfonts}
\usepackage{times,txfonts}
\usepackage{nicefrac}
\usepackage[colorlinks=true,linkcolor=blue,urlcolor=blue,citecolor=blue,pdfusetitle]{hyperref}
\usepackage{hyperref}
\usepackage[normalem]{ulem}
\usepackage{calrsfs}
\DeclareMathAlphabet{\mathcal}{OMS}{cmsy}{m}{n}
\usepackage{amssymb}
\usepackage{natbib}
\begin{document}

\title{Entanglement Classification via Neural Network Quantum States}

\author{Cillian Harney$^{1,2}$, Stefano Pirandola$^{2}$, Alessandro Ferraro$^{1}$, and Mauro Paternostro}
\affiliation{Centre for Theoretical Atomic, Molecular and Optical Physics, School of Mathematics and Physics, Queen's University Belfast, Belfast BT7 1NN, United Kingdom\\
$^{2}$Computer Science and York Centre for Quantum Technologies,University of York, York YO10 5GH, United Kingdom}

\date{\today{}}

\begin{abstract}
The task of classifying the entanglement properties of a multipartite quantum state poses a remarkable challenge  due to the exponentially increasing number of ways in which quantum systems can share quantum correlations. Tackling such challenge requires a combination of sophisticated theoretical and computational techniques. In this paper we combine machine-learning tools and the theory of quantum entanglement  to perform entanglement classification for multipartite qubit systems in pure states. We use a parameterisation of quantum systems using artificial neural networks in a restricted Boltzmann machine (RBM) architecture, known as Neural Network Quantum States (NNS), whose entanglement properties can be deduced via a constrained, reinforcement learning procedure. In this way, Separable Neural Network States (SNNS) can be used to build entanglement witnesses for any target state.
\end{abstract}

\maketitle{}


\newcommand\litem[1]{\item{#1.}}




\section{Introduction}
 As the size of a quantum system grows, the number of accessible states, and thus the Hilbert space dimension, scales exponentially. Therefore, the amount of information required for a complete description of a many-body quantum state quickly grows uncontrollably. For this reason, as exact descriptors of many-body systems becomes intractable, we should quickly turn to mathematical models for the simulation of quantum states.
 
Very recently, {machine learning} has become a prominent numerical tool for the assessment of problems of overwhelming complexity, with applications in many areas of physics~\cite{review1,review2,review3}. In particular, \textit{Artificial Neural Network} (ANN) architectures have been shown to provide excellent representations of quantum systems, due to their efficiency in dimensional reduction, and sufficient expressive power to provide efficient simulations and insight in quantum problems with high dimensional Hilbert spaces, inaccessible by many other analytical or numerical means. A key instance of problems where ANN-based approaches hold the promises for a game-changing contribution is the discrimination of entangled and separable states, which is a known NP-hard classification problem in quantum information processing~\cite{horodeckiRMP}.

In this work, we employ the recently introduced \textit{neural network quantum states} (NNSs)~\cite{Carleo1},  which are ANN architectures of the restricted Boltzmann Machine (RBM) form, to build an accurate entanglement-separability classifier that we show to be effective in  both witnessing  multipartite entangled states and identify the $k$-inseparability class of generic multipartite quantum states. Our tool requires minimum adaptation to the form of possible input states, as we show by addressing various multipartite qubit states, including linear cluster states, which are crucial resources for measurement-based for quantum computation~\cite{BriegelNatPhys}.

The remainder of this paper is organized as follows. In Sec.~\ref{NNS} we introduce the concept of NNS and their parameterization, while Sec.~\ref{class} is dedicated to our strategy for the characterization of pure multipartite entangled states. In Sec.~\ref{results} we present the results of our analysis for a series of benchmark examples including linear cluster states, while Sec.~\ref{conc} is for our conclusions and a sketch of our future directions of investigation.

\section{Neural Network States}
\label{NNS}
As mentioned above, NNSs provide a parameterisation for the wavefunction of quantum systems by means of RBM-like architectures~\cite{Carleo1}, which have recently received considerable attention~\cite{NNSoverview}. RBMs consist of a single visible and hidden layer of neurons, mediated by weighted inter-layer connections and with no intra-layer links. The visible layer embodies the physical degrees of freedom of the system, whilst the hidden one is used to distribute information across the network. The optimization of the latter is 
the intrinsic purpose of any ANN.

We consider a generic, pure NNS with $N$ discrete-valued degrees of freedom, for example a system of $N$ qubits $\vec{s} = \{s_i\}_{i=1,\ldots,N}$, as a visible layer of binary-valued neurons, fully connected to a hidden layer of $H$ hidden binary-valued neurons $\vec{h} = \{h_j\}_{j=1,\ldots,H}$, where $s_i, h_j = \{ -1,1\}$. These connections are mediated by the variational parameters of the network $\Omega = \{ a_i, b_j, \mathcal{W}_{ij} \}$. The wavefunction of this state is thus given by
\begin{equation}
\Psi_{\Omega}(\vec{s},\vec{h} )= \sum_{\vec{h}} e^{\sum_i s_i a_i + \sum_{ij} \mathcal{W}_{ij} h_j s_i + \sum_j h_j b_j}.
\end{equation}
The hidden layer of binary neurons $\vec{h}$ can be readily traced out, due to the lack of intra-layer connections, thus providing a representation depending only on $\Omega$ and the physical spin-like variables in the visible layer
\begin{gather} \label{eq:ansatz}
\Psi_{\Omega}(\vec{s}) = e^{\sum_i s_i a_i} \prod_j 2\cosh{\Big({\sum_i \mathcal{W}_{ij} s_i + b_j} \Big)}.
\end{gather}
The actual NNS can thus be written as $\ket{\Psi_\Omega} = \sum_{\vec{s}} \Psi_{\Omega}(\vec{s})\ket{\vec{s}}$ (up to an irrelevant normalisation constant). Note that this ansatz describes \textit{pure states}. 
\begin{figure}[t!]
\includegraphics[width=\columnwidth]{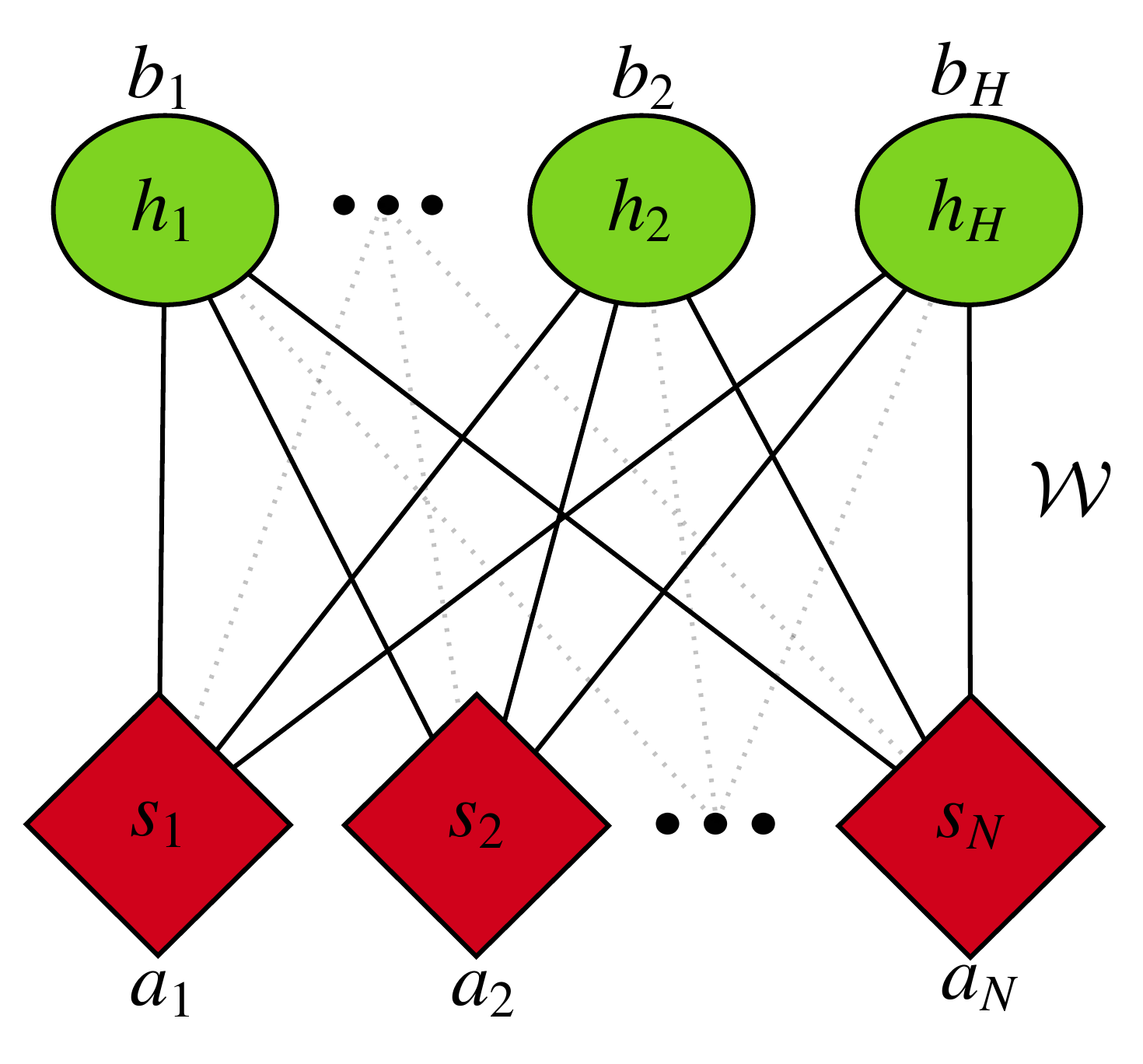}
\caption{Illustration of a $N$-qubit NNS based on a RBM of $N$ binary artificial visible neurons, and $H$ binary artificial hidden neurons used to mediate the correlations within the system. There are $N H$ weighted connections and $N + H$ total neural biases.}
\end{figure}

\section{Pure State Entanglement Classification}
\label{class}
The non-local features of the RBM architecture allows for the assessment of entanglement throughout the system. The capacity of the NNS to represent multipartite entangled state is based on the amount of network parameters being exploited~\cite{Dong}. Utilising more hidden neurons in the RBM structure increases the sets of weights and biases, and thus the usefulness of the network state. However, representing a pure state via the ansatz in Eq.~(\ref{eq:ansatz}) requires the exact parameterisation of the $N$-qubit state in terms of the neural network set of parameters $\Omega = \{ a_i, b_j, \mathcal{W}_{ij} \}$. Fortunately, NNS are constructed so that they can undergo variational evolution using a learning-optimisation procedure. Therefore it is straightforward to implement a learning scheme that variationally evolves a NNS $\ket{\Psi_{\Omega}}$ into a known target state $\ket{\varphi}$ through the maximisation of state fidelity. 

Under the assumption that any entangled state is learnable, 
it is interesting to investigate the relationship between the set of parameters entering a NNS and the separability properties of the state. 
We will see that the use of Separable Neural Network States (SNNS) in conjunction with such a fidelity-maximisation learning scheme, target states can be classified based on their entanglement properties. 

\subsection{Quantum state representation and translation}
In order to provide a systematic way of translating generic $N$-qubit states into a form represented by a NNS, we use the following approach: Given a {\it blank} NNS $\ket{\Psi_{\Omega}}$ with $N$ visible neurons, $H$ hidden neurons, parametezised by a the set $\Omega$, and given a target state $\ket{\varphi}$, we wish to optimise $\Omega$ in a way that $\ket{\Psi_{\Omega}}$ most closely approximates $\ket{\varphi}$. This can be achieved using a learning procedure that iteratively updates $\Omega= \{ a_i, b_j, \mathcal{W}_{ij} \}$ so as to achieve the set $\Omega^{\prime}$ 
for which  the fidelity between the NNS and the target state is maximum
$\forall i \in [1,N], \forall j \in [1,H]\nonumber$. As the target state is known and fixed throughout the entire optimisation, state fidelity can be computed as a multi-variable function dependent on the neural network parameters as
\begin{equation}
\mathcal{F}(\Omega) = \sqrt{
	\frac{\left\vert\braket{\Psi_{\Omega} | \varphi}\right\vert^2}{\braket{\Psi_{\Omega} | \Psi_{\Omega}} \braket{\varphi | \varphi}}}.
\end{equation}
The quantities in this expression can be computed as classical expectation values over probability distributions defined by the state at hand, which delivers a readily computable fidelity between the adaptive RBM state and the target state~\cite{CarleoBJ}
\begin{equation}
\begin{aligned}
\mathcal{F}(\Omega) &=	
	\sqrt{
	\left[ \frac{\sum_{\vec{s}}\> ({\varphi}/{\Psi_{\Omega}}){\left| \Psi_{\Omega} \right|}^2}{\sum_{\vec{s}} {\left| \Psi_{\Omega} \right|}^2} \right]
	\left[ \frac{\sum_{\vec{s}} \> ({\Psi_{\Omega}^*}/{\varphi^*}) {\left| \varphi \right|}^2}{\sum_{\vec{s}}  {\left| \varphi \right|}^2} \right]
	}\\
	&= \sqrt{
	\left \langle \frac{\varphi}{\Psi_{\Omega}} \right \rangle_{\Psi_{\Omega}}
	\left \langle \frac{\Psi_{\Omega}}{\varphi} \right \rangle_{\varphi}^*
	},	
	\end{aligned}
\end{equation}
where, generally, $\left\langle A \right\rangle_\alpha$ denotes a statistical expectation value of the quantity $A$ over the probability distribution $\alpha$. Computation of these expectation values can be achieved via standard Markov Chain Monte Carlo techniques.
This quantity can then be used to implement a learning scheme in which at every iteration of the optimisation, the network parameters are adjusted to provide a positive fidelity gradient, until convergence at a maximum (ideally unit) value is achieved. In practice, it is more convenient to consider the negative logarithm of the overlap as it possesses a more compact form of gradient, converting this maximisation into a minimisation. Defining $\mathcal{O}_i = {\partial_{\Omega_i} \ln (\Psi_\Omega)}$, the loss function and its gradients for this learning scheme are given formally as
\begin{equation}
    \begin{aligned}
\mathcal{L}(\Omega) &= -\log(\mathcal{F}(\Omega)),\\
\partial_{\Omega_i} \mathcal{L}(\Omega) &= {\langle\mathcal{O}_i^* \rangle_{\Psi_{\Omega}}} - \frac{ \left\langle ({\varphi}/{\Psi_{\Omega}}), \mathcal{O}_i^*\right\rangle_{\Psi_{\Omega}}}{\langle {\varphi}/{\Psi_{\Omega}}\rangle_{\Psi_{\Omega}}}.
\end{aligned}
\end{equation}
With this at hand, we can now construct a learning scheme using stochastic gradient descent (SGD). Updates to the $i^{\text{th}}$ network parameter of the RBM wavefunction at the $k^{\text{th}}$ iteration will be given by
\begin{equation}
\Omega_i^{k+1} = \Omega_i^k - \eta \> \partial_{\Omega_i^k} \left\langle \mathcal{L}(\Omega) \right\rangle,
\end{equation}
where $\eta$ is the learning rate of the process. 
Over enough iterations and a small enough learning rate, convergence is guaranteed, and the network variational state is optimised to reconstruct the desired target state. The latter can be generally represented as a sum of Kronecker functions with unique probability amplitudes $\varphi(\vec{s}_j) = \sum_{i = 1}^{2^N} \alpha_i \delta(s_i, s_j)$ where $\delta(s_i, s_j)$ is equal to unity if and only if $s_i = s_j$. However, the use of such Kronecker functions within the target wavefunction provide a very difficult optimisation problem for the learning procedure, due to the infinite magnitude of the gradients on the associated free-energy surface. By smoothing the target wavefunction into a sum over Gaussians, the task becomes much more manageable, while retaining accuracy if sufficiently small variances are used. The wavefunction for this approximation to the target state thus becomes 
$\varphi(\vec{s}_j) \approx \sum_{i=1}^{2^N} \alpha_i e^{-{(\beta_i - \beta_j)^2}/{\sigma^2}}$,
where $\beta_k = \text{bin}(\vec{s}_k)$ is the binary conversion of the $k^\text{th}$ $N$-qubit basis state, and $\sigma^2$ is the variance of the Gaussian packet.\par

\begin{figure}[t!]
\includegraphics[width=\columnwidth]{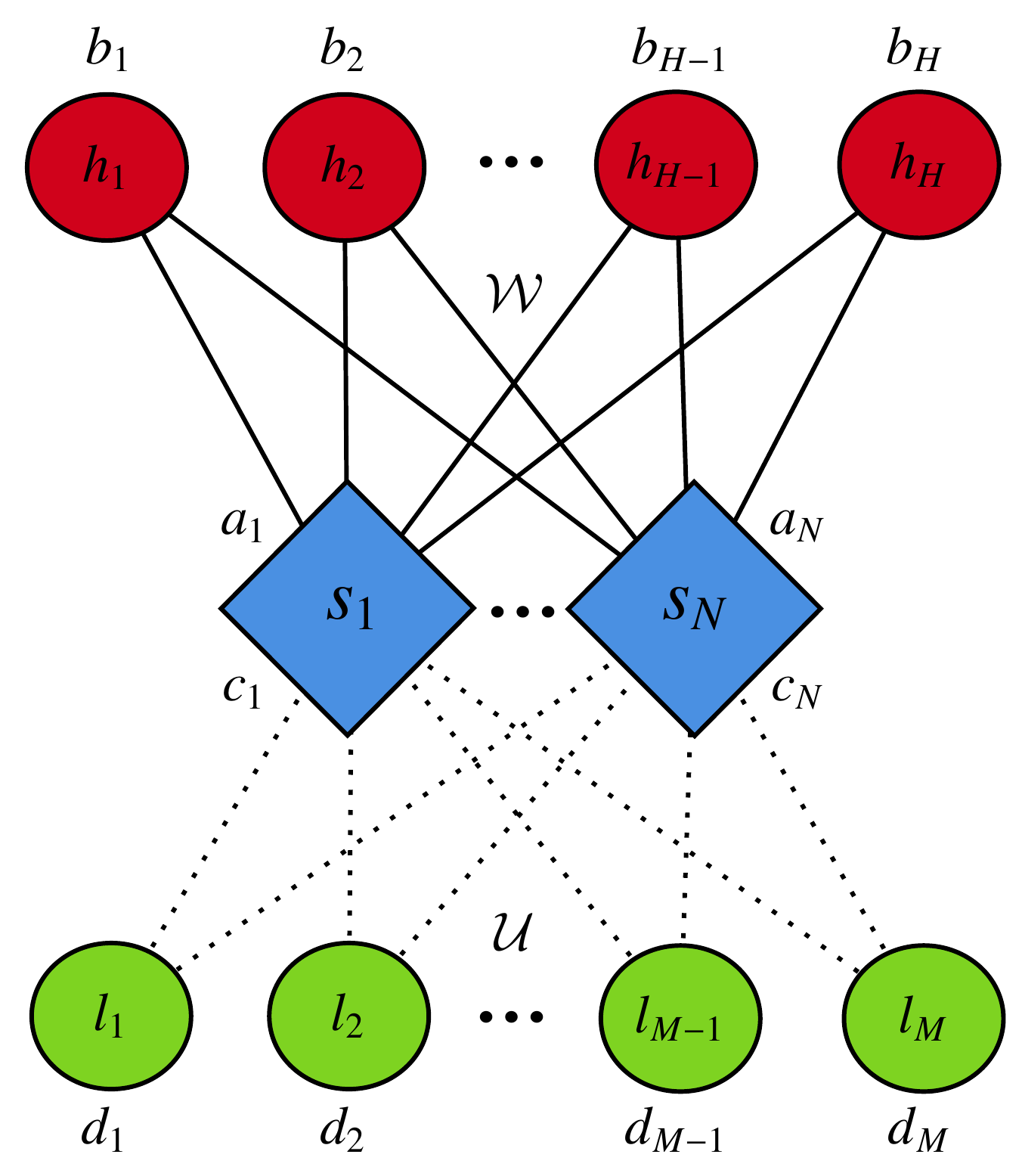}
\caption{
{Graphical depiction of a NNS describing a system of $N$ qubits, using a RBM machine of $N$ binary artificial visible neurons, $H$ binary artificial hidden neurons and $M$ binary artificial hidden neurons that mediate amplitude and phase correlations, respectively. There are $N (H + M)$ weighted connections and $2N + H+M$ total neural biases.}}
\label{Figure2}
\end{figure}

The ability of an NNS to represent local phase within a target state is dictated by the nature of the ANN parameters. A NNS with complex weights and biases 
is able to generate generally complex amplitudes such that $\Psi_{\Omega}(\vec{s}) = re^{i\varphi}$  ($r \in \mathbb{R}, \varphi \in [0,2\pi]$) for some input vector of qubit configurations. Thus, NNS with purely real parameters are only capable of simulating positive wavefunctions, up to a global phase factor.\par
However, introducing a non-trivial phase structure into a target state increases the complexity of the optimisation problem. It is instructive to instead utilise an additional layer of hidden neurons, $\vec{l} = \{l\}_{k=1,\ldots,M}$ with an associated set of weights and biases $\Xi = \{c_i,d_k,\mathcal{U}_{ik}\}$ dedicated to learning local phase factors of a target state \cite{Tomog}. Thus the original hidden layer $\vec{h}$ and its weights and biases $\Omega$ becomes the dedicated amplitude-learning parameter set (see Fig.~\ref{Figure2}). This introduces a new global NNS ansatz, combining the contribution of both layers, and reading
\begin{equation}
\label{globalNNS}
\ket{\Psi_{\Omega,\Xi}}= \sum_{\vec{s}} e^{2i\pi\>\Phi_{\Xi}(\vec{s})}\Psi_{\Omega}(\vec{s})\ket{\vec{s}}
\end{equation}
with $\Phi_{\Xi}(\vec{s}) 
\in [0,1]$.
In our numerical experiments on target-state reconstruction, the method of natural gradient descent~\cite{Sorella} was found to be more effective than that of stochastic gradient descent, ans is thus adopted in what follows. Updates to the $i^\text{th}$ parameter of the network at the $k^\text{th}$ iteration are thus given by
\begin{equation}
\label{updateXi}
\Xi_i^{k+1} = \Xi_i^k - \eta \> \sum_{j} \langle S_{ij} \rangle^{-1} f_j
\end{equation}
where $\langle S_{ij} \rangle$ denotes the elements of a covariance matrix and $f_j$ the elements of a generalised force vector, both defined in Appendix~\ref{AppA}. The updates to $\Xi_i^{k}$ in Eq.~\eqref{updateXi} are based on a natural metric of the variational subspace being explored, which greatly enhances the optimisation process.

From this point forward we omit reference to the phase learning layer unless required, but recognise that its application is synonymous with the original NNS design.

\subsection{Separable neural network states and multipartite entanglement}
\begin{figure}[t!]
\includegraphics[width=\columnwidth]{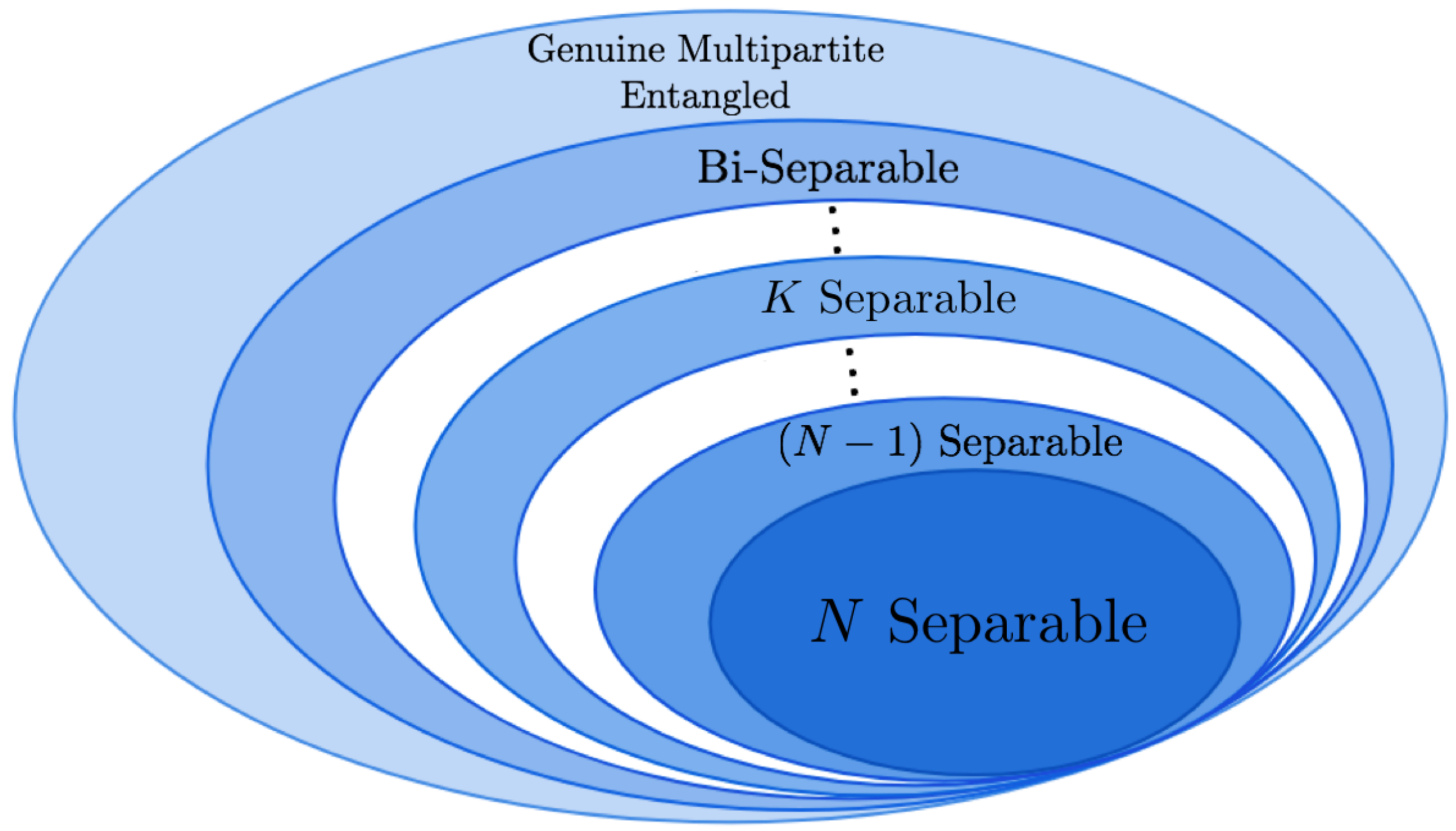}
\caption{Geometric representation of the hierarchy of multipartite entangled states for $N$-partite quantum states. For every partition of an $N$ party system, there exists a set of states that admit $K$-separability, $\mathcal{U}_{K}$. States that are $K$ separable are also representable as $(K+1)$-separable states, thus $\mathcal{U}_{K} \subset \mathcal{U}_{K+1} \subseteq \mathcal{U}$ (where $\mathcal{U}$ is the total set of states). If a state $\rho \in \mathcal{U}$ but $\rho \not\in \mathcal{U}_{K}, \forall K  \in [2,N] $, then $\rho$ is genuinely multipartite entangled \cite{hier}.}
\label{FigureEnt}
\end{figure}
The ability to effectively enforce properties of separability onto a NNS is extremely useful and integral to the entanglement classification protocol addressed in this paper. In order to enforce a particular form of separability into a pure multipartite state, we must first determine the number of ways in which it may possess entanglement. An $N$-qubit pure quantum state is said to be $K$-separable if it is the tensor product of $K=2,..,N$ parties of the total system, where $N$-separability coincides with full separability. Differently, if a state is genuinely multipartite entangled it cannot be factorised into any tensor product representation ($K=1$).

We define a pure $K$-separable state
$\ket{\Psi} = \bigotimes_{i=1}^K \ket{\psi_{\mathcal{S}_i}}$,
such that $\mathcal{S} = \{\mathcal{S}_i\}_{i=1,\ldots,K}$ is a set of $K$ disjoint subsets of the $N$ parties of the total quantum system, i.e. $\mathcal{S}_i \cap \mathcal{S}_j = \emptyset,\>\> \forall i,j \in \{1,\ldots,K\}$.\par
However, at the level of constructing specific separability sets according to $\mathcal{S} = \{\mathcal{S}_i\}_{i=1,\ldots,K}$, the number of ways a $K$-separable state can be invoked is highly degenerate (cf. Appendix~\ref{AppB}). 
Therefore, one can describe a state 
as $K_j$-separable (i.e. a $K$-subseparability) (with $j \in [1,P_K]$ where $P_K$ is the degeneracy of $K$ partitioning) in order to specify the exact form of $K$-separability being addressed.

We are thus left to deduce a translation of $K_j$-separability from its fundamental definition into a set of relations for the parameter set of the NNS. Given a set of partitions $\{\mathcal{S}_m\}_{m=1,\ldots,K}$ we wish to find the network conditions such that the NNS with $N$ visible neurons and $H$ hidden neurons can only reproduce states with such form of separability. This can be achieved by solving the following equation
\begin{equation}
\begin{aligned}
\Psi_{\Omega}(\vec{s}) &= e^{\sum_i s_i a_i} \prod_{j=1}^H \prod_{k=1}^K \psi_{\mathcal{S}_k,j} \\
 &=e^{\sum_i s_i a_i} \prod_{j=1}^H 2\cosh{\Big({\sum_i \mathcal{W}_{ij} s_i + b_j} \Big)},
\end{aligned}
\end{equation}
where $\psi_{\mathcal{S}_k}$ is an ansatz for a ``local" wavefunction for each collection of entangled qubits. In this way we are requesting that $\Psi_{\Omega}(\vec{s})$ takes a desired product form, and the required conditions can thus be derived from of the solutions of 
\begin{equation}
\label{separcond}
\prod_{j=1}^H \prod_{k=1}^K \psi_{\mathcal{S}_k,j}  = \prod_{j=1}^H 2\cosh{\Big({\sum_i \mathcal{W}_{ij} s_i + b_j} \Big)}
\end{equation}
since identical products are taken over the hidden neurons in both cases. The goal of this task is to transform the right-hand side (RHS) of Eq. (\ref{separcond}), currently capable of describing all forms of separable states, into a form that aligns with the left-hand side (LHS) and therefore the separability properties of the state.

This separation can be achieved by performing segmentations of the neural network architecture according to the separability being imposed. Each set of potentially entangled qubits $\mathcal{S}_m$ is fully connected to a dedicated set of hidden neurons $\mathcal{H}_m$ ($\mathcal{W}_{i\in\mathcal{S}_m,j\in\mathcal{H}_m} \neq 0$), but are fully \textit{disconnected} to all other hidden neurons ($\mathcal{W}_{i\in\mathcal{S}_m,j\not\in\mathcal{H}_m} = 0$). Thus, there exist $K$ disjoint sets of hidden neurons $\{\mathcal{H}_m\}_{i=1,\ldots,K}$ corresponding to $K$ disjoint sets of qubits $\{\mathcal{S}_m\}_{m=1,\ldots,K}$. Performing this segmentation, the RHS of Eq. (\ref{separcond}) becomes
\begin{equation}
\prod_{m=1}^K\prod_{j\in\mathcal{H}_m} 2\cosh{\Big({\sum_{i\in\mathcal{S}_m}  \mathcal{W}_{mj} s_i + b_j} \Big)},
\end{equation}
In this way, an $N$-qubit $K_j$-Separable Neural Network State (SNNS) is defined as an RBM with $N$ visible neurons segmented into disjoint sets $\{\mathcal{S}_m\}_{m=1,\ldots,K}$ and $H$ hidden neurons segmented into disjoint sets $\{\mathcal{H}_m\}_{m=1,\ldots,K}$ mediated by complex variational parameters $\Omega = \big\{ a_i,b_j,\mathcal{W}_{ij} \big\}$ with the property
\begin{equation}
\label{netcond}
\mathcal{W}_{i\in\mathcal{S}_m,j\not\in{\mathcal{H}_m}} = 0,\qquad \forall
\begin{cases} 
i\in [0,N],&\\ 
j\in[0, H],& \\
m\in[0,K].&
\end{cases}
\end{equation}
\begin{figure}[t!]
\includegraphics[width=\columnwidth]{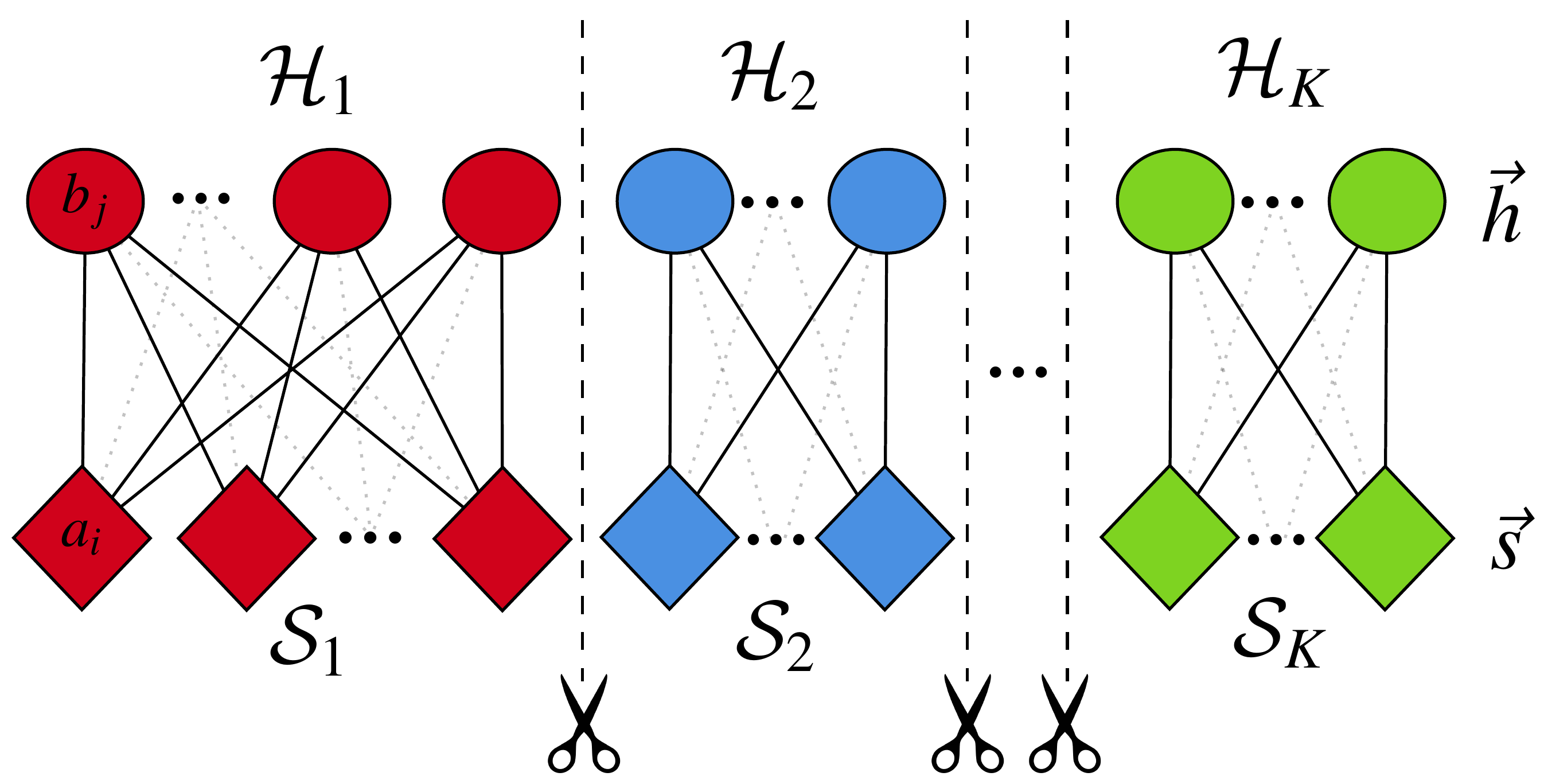}
\caption{Graphical depiction of a $N$-qubit $K_j$-SNNS based on a RBM of $N$ binary artificial visible neurons and $H$ binary artificial hidden neurons that mediate the correlations within the system. Separability under the set of partitions $\{\mathcal{S}_m\}_{m=1,\ldots,K}$ is enforced by performing segmentations throughout the network. Each set of qubits $\mathcal{S}_m$ is fully connected to a set of hidden neurons $\mathcal{H}_m$ ($|\mathcal{H}_m| \neq |\mathcal{S}_m|$ generally) but is independent from all other hidden neurons. There are $\sum_{m=1}^K |\mathcal{H}_m| |\mathcal{S}_m|$ non-zero weighted connections, and $N+H$ neural biases.}
\end{figure}
Such separable network architecture generally relies on a larger number of hidden neurons than that of a conventional NNS, due to the need for dedicated sets of neurons for each separable subsystem of the $N$-qubit set. However, it is important to recognise that this increase in hidden neurons \textit{does not} decrease the efficiency of the optimisation procedure. This is because the null weights in $\mathcal{W}_{ij}$ (disconnections) do not require updates during the learning protocol, and can thus be ignored. Hence the number of meaningful parameters in $\mathcal{W}_{ij}$ is given by
\begin{equation}
|\Omega_{\text{Sep}}| = N + H + \sum_{m=1}^K |\mathcal{H}_m|  |\mathcal{S}_m|,
\end{equation}
which is comparable with the number of parameters $|\Omega_{\text{Free}}| = N + H + N H$ in a free learner. This returns the computational complexity of the learning regime for SNNS to that of a typical quantum state reconstruction.\subsection{Entanglement Classification}
With the necessary tools in place, an entanglement classification protocol can be devised. The maximum fidelity learning regime allows a blank, randomised NNS to undergo variational evolution in order to converge towards a pre-defined pure quantum state, and accurately simulate the target one. Such learning process operates under the assumption that the target state is simulatable by the the NNS being utilised. If a typical NNS is used then this is true, as there are no restrictions/rule on the values the network parameters can take. In this way we call generic NNS \textit{free learners} of target states.

However, if one uses a $K$-separable SNNS, this is not necessarily the case. A $K$-separable SNNS is a neural network state that possesses network parameters in accordance with Eq. (\ref{netcond}) which can only simulate states with this form of $K$-separability. Therefore, if a $K$-separable SNNS (or \textit{restricted learner}) is used in conjunction with the maximum fidelity learning scheme in order to reconstruct a pure state $\ket{\varphi}$, it will only achieve maximum fidelity if the target state is also $K$-separable. Otherwise, if the target state is in fact $K^\prime$-separable the SNNS will learn to optimise its fidelity to the maximum value that a $K$-separable state can achieve with the $K^\prime$-separable state $\ket{\varphi}$. If $K^\prime$-separability is a higher order of entanglement than $K$, then the optimised fidelity will be a value less that unity. Yet if $K^\prime$-separability is a lower order of entanglement, then it will be representable as $K$-separable also.\par
The classification protocol thus follows: For a pure $N$-partite quantum state $\ket{\varphi}$ which can be optimally reconstructed via unrestricted maximal fidelity learning, if a $K_j$-separable SNNS ${\ket{\Psi_{\Omega}^{K_j}}}$ (defined by the set of disjoint sets of qubits $\{\mathcal{S}_i\}_{i=1,\ldots,K}$) is unable to reconstruct $\ket{\varphi}$ through the same optimisation scheme
\begin{equation}
\begin{aligned}
{\ket{\Psi_{\Omega}^{\text{Free}}}} &\xrightarrow[\text{optimise}]{} \ket{\Psi_{\Omega^\prime}^\text{Free}} \equiv \ket{\varphi},\\
{\ket{\Psi_{\Omega}^{{K_j}}}} &\xrightarrow[\text{optimise}]{} {\ket{\Psi_{\Omega^\prime}^{{K_j}}}} \equiv \ket{\varphi^\prime} \not\equiv \ket{\varphi},
\end{aligned}
\end{equation}
then the target state $\ket{\varphi}$ must possess entanglement within at least one of the partitions $\mathcal{S}_i$. This approach provides valid separability criteria for conclusive classification of pure quantum states.

\subsection{Witnesses and measures}
The construction of a reliable and consistent entanglement classification procedure requires a quantifiable measure of performance for NNS target state reconstructions. As ANN learning is numerical in nature, it may be prone to statistical errors and possible flaws due to the size of Hilbert space being explored, and infinite possible variational updates that can be made. Hence, we resort to statistics to combat this.

Consider a classification protocol which uses a restricted learner $\ket{\Psi_\Omega^K}$ in order to classify this form of entanglement for a target state $\ket{\varphi}$, achieving a set of fidelities $\{F_K^i\}_{i=1,\dots,M}$ over all the learning operations. In order to build a level of reliability and confidence, this protocol is performed $M$ times so to calculate an average fidelity and their variance $\langle F_K \rangle = \sum_{i=1}^M F_K^i/M$, $(\Delta F_K)^2 = \langle F^2_K\rangle - \left\langle F_K\right\rangle^2$. Given enough samples $M$ and enough hidden neurons (and thus free parameters) to ensure sufficient expressive power, we can define a \textit{performance set} $\>\mathcal{F}_{K} = \big[ \langle{F}_K \rangle - |\Delta F_K |, \langle{F}_K \rangle + |\Delta F_K |\big]$ that describes a window of reliability in the particular, separable learner being employed. Doing so equivalently for the free learner $\ket{\Psi_\Omega^\text{{Free}}}$ is also extremely important, providing a benchmark for the performance of the NNS without entanglement property restrictions. In fact, the learning performance of any restricted learner can be expressed \textit{relative} to the behaviour of the free learner, and provides a systematic method for conclusive entanglement classification. In general there are two cases in doing this
\begin{itemize}
\item $ \mathcal{F}_K \cap \mathcal{F}_\text{Free} \neq \emptyset$. In this case there is an intersection between the computed fidelity of the restricted learner and the free learner, meaning that we can classify this state as possessing entanglement properties according to this form of $K$-separability.
\item $ \mathcal{F}_K \cap \mathcal{F}_\text{Free} = \emptyset$. In this case there is no intersection between the computed fidelity of the restricted learner and the free learner. In this case the learner has only been able to reconstruct (ideally) the closest state to $\ket{\varphi}$ that possesses entanglement properties according to $\ket{\Psi_\Omega^K}$.
\end{itemize}
This approach provides a consistent rule for deciding how a target state is entangled. The application of this method relies on the accuracy of the learning regime to maintain a low variance throughout its full spectrum of fidelities, since an arbitrarily large variance will render the result redundant. Nonetheless, this method of classifying the entanglement properties of target states resembles that of \text{entanglement witnesses}. If a $K$-separable learner achieves an optimal reconstruction fidelity with a performance similar to the free learner, then the NNS \textit{witnesses} this state as entangled in this way. Otherwise, it does not witness the state and this binary classification delivers the contrary result.\par
A much more detailed classification can be carried out by more closely investigating the resultant fidelities of all restricted learners according to a set of separabilities, not just those that achieve optimal fidelities with respect to the free learner. Instead, one can consider the relative fidelity of the $K_j$-separable learner with respect to the free learner as a measure of how much $K_j$-separability is manifested within the target state. Defining the relative fidelity
\begin{equation}
\mathcal{R}_{K_j} = \frac{\langle{F_{K_j}}\rangle}{\langle{F_\text{Free}}\rangle},
\end{equation}
an approximate, local entanglement measure can be devised in accordance with the general properties of an entanglement measure \cite{VedralMeasure}
\begin{equation}
\label{entmeas}
\mathcal{E}_{K_j} = 1 - \mathcal{R}_{K_j}^2.
\end{equation}
\begin{figure}[t!] 
{{\bf (a)}\hskip3.5cm{\bf (b)}}
        \includegraphics[width=\columnwidth]{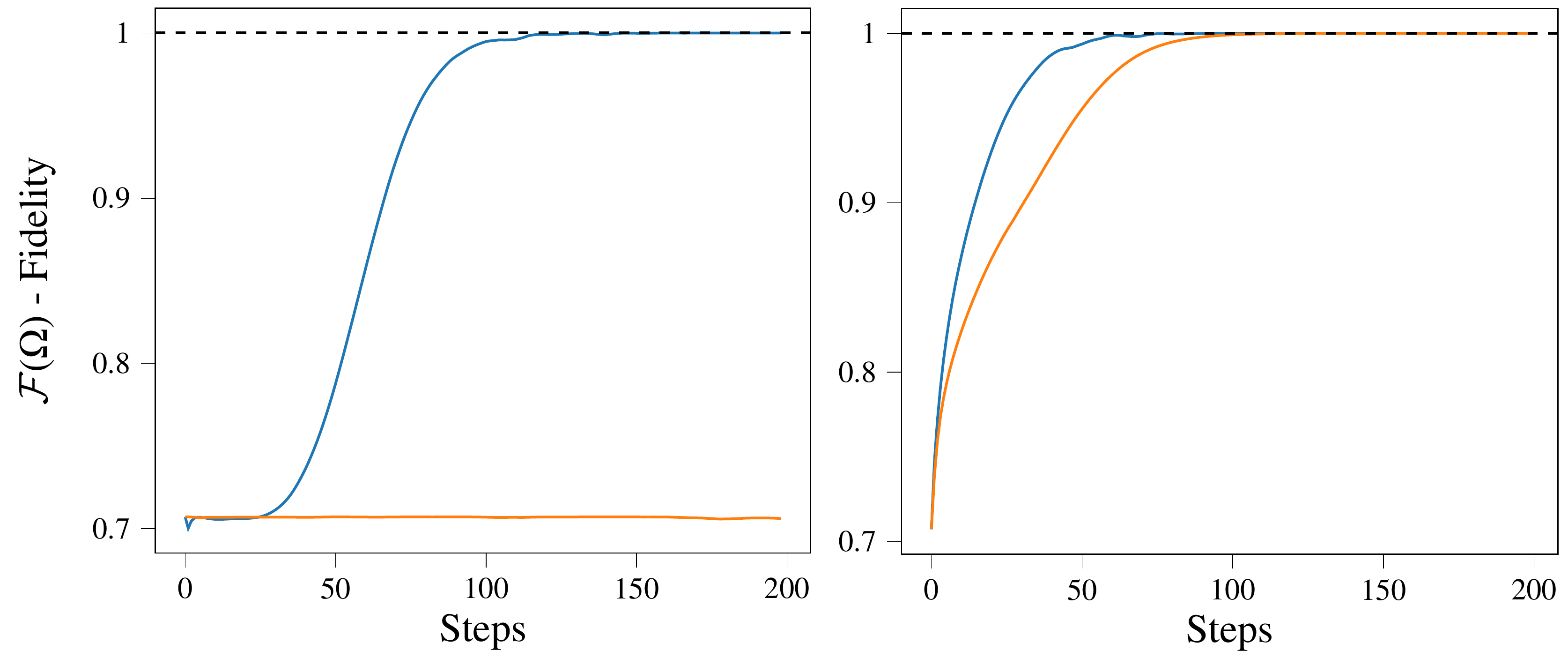}
    \caption{Classification of two-qubit quantum states. Panel {\bf (a)} reports the learning paths of a free learner $\ket{\Psi_\Omega^\text{Free}}$ (blue) and a separable learner (orange) attempting to reconstruct a two qubit Bell state $\ket{\Psi^+}=(\ket{01}+\ket{10})/\sqrt2$ such that the separable learner $\ket{\Psi_{\Omega}^{1|2}}$ is unable to acquire a maximal fidelity, whilst the free learner easily achieves unit fidelity. Instead, $\ket{\Psi_{\Omega}^{1|2}}$ converges to the maximum fidelity that the set of separable states can acquire with $\ket{\Psi^+}$, which is ${1}/{\sqrt{2}}$. Panel {\bf (b)} illustrates the behaviour of both the entangled learner  $\ket{\Psi_\Omega^\text{Free}}$ and the separable learner $\ket{\Psi_{\Omega}^{1|2}}$  converging to unit fidelity while reconstructing a separable two qubit state $\ket{\varphi} = \ket{+}_1\ket{0}_2$. }
    \label{Figure4}
\end{figure}
Note that we now refer to $K_j$-separability, such that this is a sub-genre of the more general $K$-separability. 
Classification does not necessarily require this distinction, but measurement of $K_j$-separability is not a complete reflection of $K$-separability, as there exist $P_K-1$ other contributions to this measure. Thus, a complete measure of $K$-separability requires an analysis of all such contributions.

The quantifier in Eq.~(\ref{entmeas}) is inspired from the well known Geometric Measure of Entanglement (GME)~\cite{GeoE}. 
The quantity $\mathcal{E}_{K_j}$ strives at quantifying the lack of representability according to $K_j$-separability. In an ideal scenario, all optimisation procedures are perfectly convergent such that for a target state $\ket{\varphi}$, any SNNS $\ket{\Psi_\Omega^{K_j}}$ will reconstruct the closest $K_j$-separable state to the target state $\ket{{\varphi_{K_j}^\prime}}$. One can define the overlap between such states as the \textit{critical fidelity} 
\begin{equation}
\alpha_{K_j} = \max_{K_j \text{-sep}}{\lVert \braket{\varphi | {\varphi_{K_j}^\prime}} \rVert},
\end{equation}
as it defines maximum fidelity between a target state and the set of $K_j$-separable states. In such ideal case, the relative fidelity and local entanglement measure become
\begin{equation}
\begin{aligned}
\mathcal{R}_{K_j} &= {\frac{ | \braket{\varphi | {\varphi_{K_j}^\prime}}  |}{\sqrt{\braket{{\varphi_{K_j}^\prime}| {\varphi_{K_j}^\prime}} \braket{\varphi | \varphi}}}} = \lVert \braket{\varphi| {\varphi_{K_j}^\prime}} \rVert 
\end{aligned}
\end{equation}
so that $\mathcal{R}_{K_j}$ recovers the critical fidelity and $\mathcal{E}_{K_j}$ the GME for multipartite, pure state entanglement.

 \begin{figure*}[t!] 
{\hskip1.5cm\bf (a)}\hskip5cm{\bf (b)}\hskip5cm{\bf (c)}
\includegraphics[width=2\columnwidth]{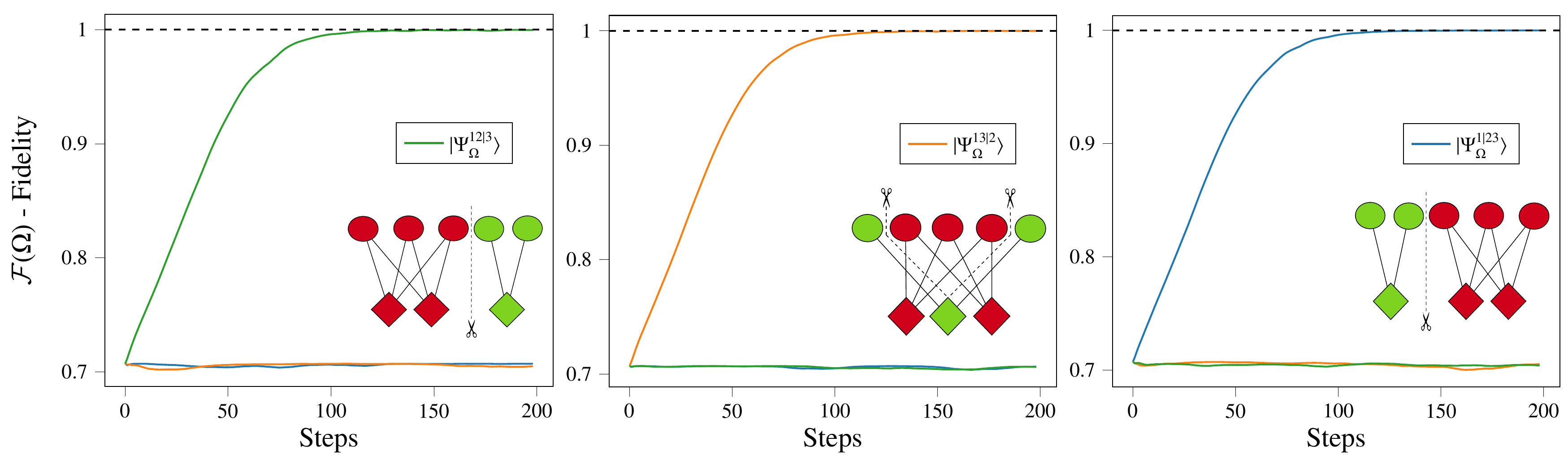}
    \caption{Examples of the ability of SNNS to distinguish between $K_j$-separable learners. Here, three tri-separable learners of the forms $\ket{\Psi_\Omega^{12|3}}$ ({green}), $\ket{\Psi_\Omega^{13|2}}$ ({orange}) and $\ket{\Psi_\Omega^{1|23}}$ ({blue}) are employed to distinguish between the exact form of separability of target state. In each case, only the learner with the correct form of tri-separability can conclusively classify the target state through convergent learning. In panel {\bf (a)} we have considered $\ket{\varphi} = \ket{\Phi^+}_{\{1,2\}}\ket{+}_3 $, in panel {\bf (b)} $\ket{\varphi} = \ket{\Phi^+}_{\{1,3\}}\ket{+}_2$, while panel {\bf (c)} is for $\ket{\varphi} = \ket{+}_1\ket{\Phi^+}_{\{2,3\}}$. }
    \label{Figure5}
\end{figure*}

\begin{figure*}[t!] 
\hskip1.5cm{\bf (a)}\hskip5cm{\bf (b)}\hskip5cm{\bf (c)}
\includegraphics[width=2\columnwidth]{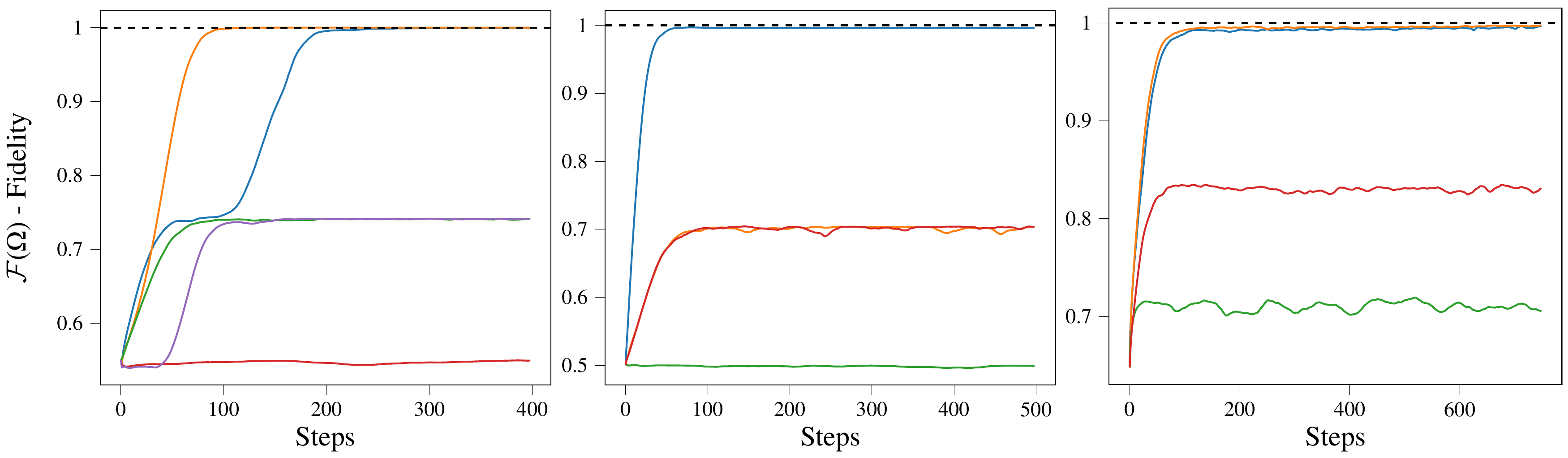}
    \caption{Entanglement witnessing via NNS. Panel {\bf (a)} depicts a four-qubit ``double Bell State", i.e. a product state of two Bell states in the $\{1,2\}$-vs-$\{3,4\}$ bipartition. We show results gathered by employing learners prepared in the states $\ket{\Psi_\Omega^{\text{Free}}}$ ({orange curve}), $\ket{\Psi_\Omega^{12|34}}$ ({blue curve}),   
       $\ket{\Psi_\Omega^{12|3|4}}$ ({green curve}), 
	$\ket{\Psi_\Omega^{1|2|34}}$ ({purple curve}),
	 $\ket{\Psi_\Omega^{1|2|3|4}}$ ({red curve}).
         The NNS with in the appropriate separable form (blue curve) achieves maximal fidelity throughout the optimisation process, whilst the triseparable and fully separable learners achieve sub-optimal convergences.
         Panel {\bf (b)} depicts a similar situation but with a biseparable state in the splitting $\{1,2,3\}$-vs-$\{4\}$ containing a tripartite entangled $\ket{\text{GHZ}}$ state, with a set of learners in the separable forms $\ket{\Psi_\Omega^{123|4}}$ ({blue}),   
       $\ket{\Psi_\Omega^{12|34}}$ ({orange}), 
	$\ket{\Psi_\Omega^{1|2|34}}$ ({green}),
	 $\ket{\Psi_\Omega^{12|3|4}}$ ({red}).
The NNS in the appropriate separable form (blue line) achieves maximal fidelity throughout the optimisation process, whilst the triseparable learners achieve suboptimal convergences. Panel {\bf (c)} reports a similar classification process for a random six-qubit state that is biseparable in the $\{1,2,3\}$-vs-$\{4,5,6\}$ bipartition using $\ket{\Psi_\Omega^{\text{Free}}}$ ({blue}),   
       $\ket{\Psi_\Omega^{123|456}}$ ({orange}), 
	$\ket{\Psi_\Omega^{1|3|2456}}$ ({red}),
	 $\ket{\Psi_\Omega^{1|2|3|4|5|6}}$ ({green}).   Panel {\bf (a)} is for $\ket{\varphi} = \ket{\Phi^+}_{\{1,2\}} \ket{\Phi^+}_{\{2,3\}}$, panel {\bf (b)} for $\ket{\varphi} = \ket{\text{GHZ}}_{\{1,2,3\}}\ket{+}_4$ and  panel {\bf (c)} was a $\{1,2,3\}\otimes\{4,5,6\}$ biseparable 6 qubit state.}
	 \label{Figure6}
\end{figure*}

\section{Results}
\label{results}
The following results and simulations are used to illustrate the effectiveness of the entanglement classification protocol. We begin with the simplest classification problem of bipartite separability and proceed to more complex states of up to six qubits. Such sizes allow for the investigation of non-trivial multipartite entangled systems without the complications entailed by large many-body systems. Each classification is characterised by a ``learning path" which depicts the evolution of the fidelity of the free learner and a separable learner throughout the state reconstruction procedure. Learning paths which follow a convergent trajectory towards unit fidelity indicate a correct classification of the state with the separability properties of the learner in question. Paths which converge with sub-optimal fidelities, or which do not converge at all, indicate a witnessing of entanglement with respect to the appropriate form of separability.\par

We start addressing the two-qubit case, a situation for which entanglement classification is a binary decision problem as a pure two-qubit state is simply either entangled or fully separable. Fig.~\ref{Figure4} illustrates the use of SNNS for classification purposes when the target state is either the Bell state $\ket{\Psi^+}=(\ket{01}_{12}+\ket{10}_{12})/\sqrt2$ or the separable state $\ket{\varphi}=\ket{+}_1\ket{0}_2$ with $\sigma_x\ket{+}=\ket{+}$ and $\sigma_x$ the $x$ Pauli matrix. The learning paths of the SNNS performs the classification successfully, whilst the free, entangled learner learns both states with ease. Note that SNNS when targeting the Bell state achieves a fidelity of $\mathcal{F}_{{1|2}} \approx {1}/{\sqrt{2}}$ which aligns with the maximum overlap between any Bell state and the set of all separable bipartite states.\par
Increasing the target system size to three qubits immediately increases the complexity of the classification problem, such that a state is tripartite entangled, biseparable (which is three-fold degenerate) or fully separable. The degeneracy of biseparability is due to the arrangement of entanglement between parties, which the appropriate SNNS are able to distinguish. Fig.~\ref{Figure5} displays the ability of SNNS to both detect $K$-separability and identify the particular permutation of entangled parties ($K_j$-separability). A similar investigation is illustrated for the four/six qubit cases in Fig.~\ref{Figure6}, which show the power of the classification protocol that is capable of providing complete entanglement descriptions of pure target states.

\begin{figure*}[t!]
{\bf (a)}\hskip5cm{\bf (b)}\hskip5cm{\bf (c)}
\includegraphics[width=2\columnwidth]{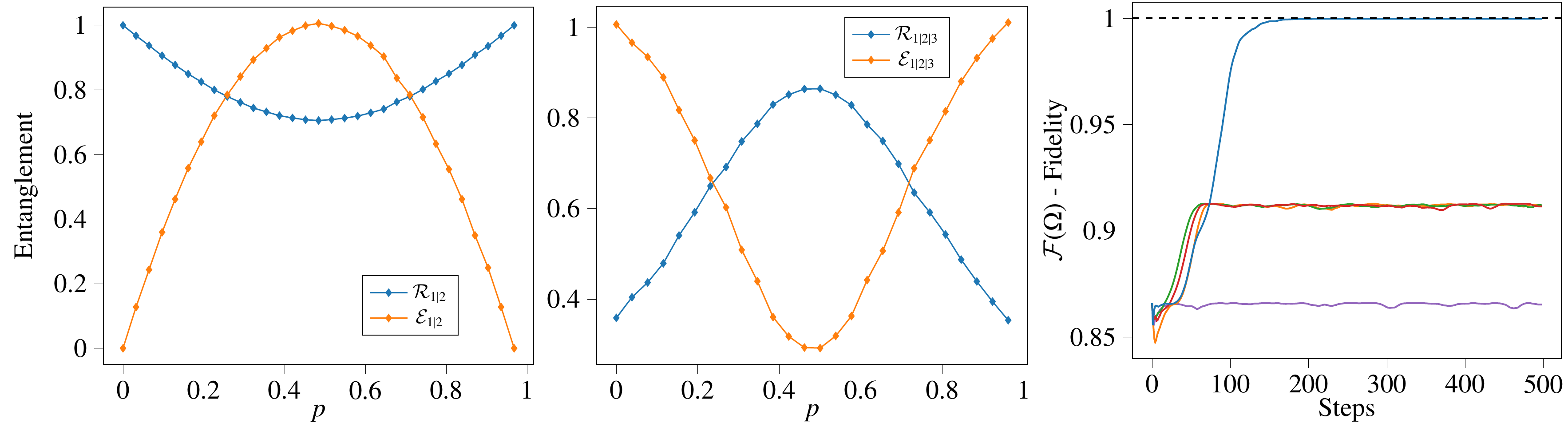}
    \caption{ Panel {\bf(a)} shows the approximate GME plotted for the two-qubit state $\ket{\varphi(p)} = \sqrt{p}\ket{00} + \sqrt{1-p}\ket{11}$. The {orange} curve plots the fully separable Relative Fidelity $\mathcal{R}_{1|2}$, whilst the {blue} curve plots the GME $\mathcal{E}_{1|2}$. 
      Panel {\bf(b)} is for the approximate geometric measure of entanglement $\mathcal{E}_{1|2|3}$ ({orange}) and relative fidelity $\mathcal{R}_{1|2|3}$ ({blue}) for the state $\ket{\varphi(p)}$ as a function of $p$. Panel (c) depicts a classification protocol aimed at the target state $\ket{\varphi(p={1/2})}$ by employing the learners 
       $\ket{\Psi_\Omega^{\text{Free}}}$ ({blue}),   
       $\ket{\Psi_\Omega^{12|3}}$ ({orange}), 
	$\ket{\Psi_\Omega^{13|2}}$ ({green}),
	 $\ket{\Psi_\Omega^{23|1}}$ ({red}) and 
	  $\ket{\Psi_\Omega^{1|2|3}}$ ({purple}), which deduce that the state is still multipartite entangled but to a lesser degree. Panel {\bf (b)} is for $\mathcal{E}_{1|2|3}$ for $\ket{\varphi(p)} = p\ket{\emph{W}} + \sqrt{1-p}\ket{\bar{\emph{W}}}$, while panel {\bf (c)} for the choice $p={1/2}$.
        }
        \label{Figure7}
\end{figure*}

Furthermore GMEs can be created to compliment the classification process and provide better insight into the entanglement properties of a target state. The example target states in Figs.~\ref{Figure4} {\bf (a)} and {\bf (b)} convey the extreme cases of maximal entanglement and complete separability respectively, however a bipartite state can contain any amount of entanglement such that its classification is less obvious. In this way, Fig.~\ref{Figure7} {\bf (a)} depicts the relative fidelity and GME of a variable Bell state $\ket{\varphi(p)} = \sqrt{p}\ket{00} + \sqrt{1-p}\ket{11}$ constructed by monitoring the performance of the separable learner with $\ket{\varphi(p)}$ for many values of $p$ in the interval $p\in[0,1]$. Similarly considering a variable, three qubit state $\ket{\varphi(p)} = p\ket{\emph{W}} + \sqrt{1-p}\ket{\bar{\emph{W}}}$, it is by no means trivial to ask whether this state is separable for any value of $p$. Constructing an approximate GME for any form of entanglement, as seen in Fig.~\ref{Figure7} {\bf (b)}, shows that $\ket{\varphi(p)}$ possesses a degree of entanglement for all values of $p\in[0,1]$, and is never fully separable. Similar investigations could be performed to measure biseparability throughout the interval and this concept may extend to any form of separability of interest in an $N$ qubit state (provided stable, convergent learning).

\begin{figure}[t!] 
      \includegraphics[width=\columnwidth]{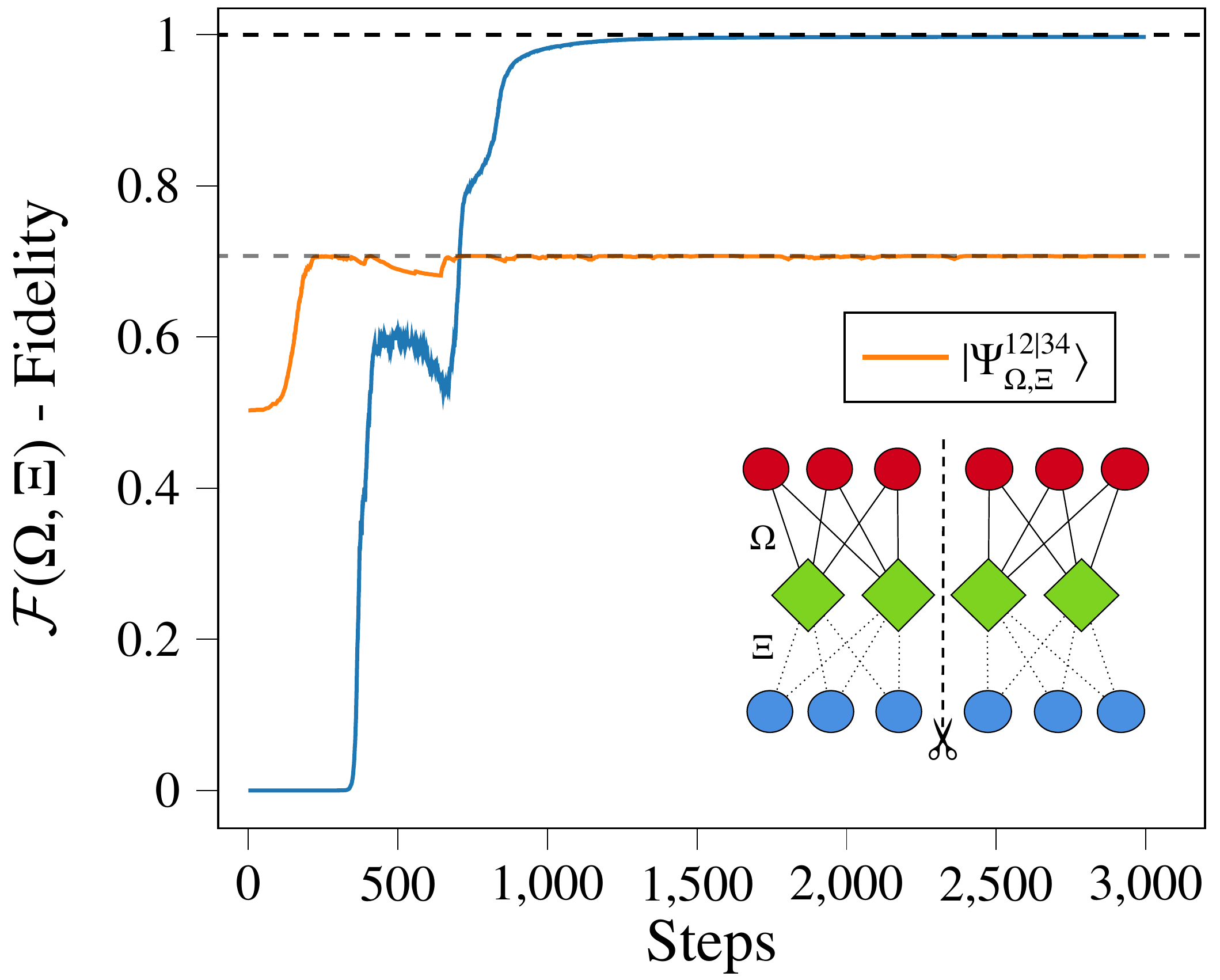}
    \caption{Entanglement classification for a four-qubit cluster state $\ket{C_4}$ built on a one-dimensional lattice, using an SNNS with a double hidden layer architecture, separating phase and amplitude learning. The free learner $\ket{\Psi_{\Omega,\Xi}^\text{Free}}$ (blue) reconstructs $\ket{C_4}$ to the maximal degree over a sufficient number of learning iterations, whilst the biseparable learner $\ket{\Psi_{\Omega,\Xi}^{12|34}}$ (orange) reaches a maximal value of fidelity of ${1/\sqrt{2}}$.
     }
    \label{Figure9}
\end{figure}

Finally, on the investigation of entangled states with non-trivial phase structures, one can utilise SNNS with network architectures according to Sec.~\ref{class}, with the applied separability conditions applied, and a natural gradient descent optimisation protocol. Particularly interesting states that fall into this category are that of cluster states, which are of great significance to quantum computing \cite{BriegelNatPhys,Cluster}. Given a $d$-dimensional square lattice of vertices $V = \{1,\ldots,N\}$, with connections between sites that define a neighbourhood, $\mathcal{N}$ connecting vertices $(i,j)$, a cluster state is given by
\begin{equation}
\ket{C_{d,\mathcal{N}}} = \prod_{(i,j)\in\mathcal{N}} U^{(i,j)} \bigotimes_{k\in{V}} \ket{+}_k,
\end{equation}
where $U^{(i,j)} = \text{diag}(1,1,1,-1)$ is a controlled-phase gate between qubits on sites $(i,j)$. The four-qubit cluster state built on a one-dimensional lattice $\ket{C_4} = (\ket{0000} + \ket{0011} +\ket{1100} - \ket{1111})_{1234}/2$ can be recast into the form $\ket{C_4} = \ket{00}_{\{12\}}\ket{\Phi^+}_{\{34\}} + \ket{11}_{\{12\}}\ket{\Phi^-}_{\{34\}}$ with $\ket{\Phi^-}=(\openone\otimes\sigma_z)\ket{\Phi^+}$. The local-phase difference between $\ket{\Phi^+}$ and $\ket{\Phi^+}$ 
removes the the $\{1,2\}\otimes\{3,4\}$ biseparability seen in the double Bell state $\ket{\Phi^+}\otimes\ket{\Phi^+}$, investigated in Fig.~\ref{Figure6} and produces the behaviour witnessed via SNNS in Fig.~\ref{Figure9}.

\section{Conclusions and further look}
\label{conc}
SNNSs offer a powerful and versatile tool in order to attack the problem of entanglement classification. With a sufficiently powerful learning mechanism, this classification method could be far reaching and proven powerful for larger, many body quantum systems.

Nonetheless, the problem of entanglement classification will remain a considerable roadblock. For systems of large $N$, the number of ways in which a state may be entangled is overwhelmingly large, and thus demanding a complete, global search of separability properties by using all possible $K$-separable learners is unfeasible. The need for some \textit{a priori} knowledge about the state, or about the form of separability one wishes to classify (such as full separability, or genuine $N$-partite entanglement) becomes essential, and greatly narrows the search. In this way, classification of larger systems becomes much more realistic using the SNNS method.

There are many further extensions and investigations worth pursuing following the introduction of SNNSs to classify entanglement. Most importantly is the extension from pure states to mixed states in a manner that maintains the power and efficiency that motivates this approach. Efficient ANN parameterisations of mixed states have been developed through the addition of a hidden ``mixing" layer to the RBM architecture to create a Neural Density Matrix~\cite{Melko,Carleo2}. Research into encoding separability properties into these machine architectures is worth exploring. An initial starting point may be aimed at generalising network conditions that invoke $K_j$-separability into $K$-separability, which could improve both pure and mixed state simulation abilities. A further, exciting, extension of this research may be to introduce generative neural network models that can numerically simulate higher dimensional quantum systems i.e qudits, and even infinite dimensional systems by constructing models within a finite-dimensional phase space. Reworking the neural network framework in a way that allows for this versatility, whilst maintaining the ability to manufacture properties such as separability and potentially Gaussianity, could provide a worthy tool that has far reaching applications in quantum communications, computing and more. 

With growing interest being accrued at the interface of quantum information and machine learning~\cite{Tomog,NISQ}, the integration of entanglement classification protocols offers an exciting avenue of exploration. 

\acknowledgements
We acknowledge support from the H2020-FETOPEN-2018-2020 TEQ (grant nr. 766900), the DfE-SFI Investigator Programme (grant 15/IA/2864), COST Action CA15220, the Royal Society Wolfson Research Fellowship (RSWF\textbackslash R3\textbackslash183013), the Leverhulme Trust Research Project Grant (grant nr.~RGP-2018-266), the EPSRC Quantum Communications Hub (grant nr. EP/T001011/1)

\bibliography{ThesisBib}

\appendix
\section{Natural Gradient Descent for Quantum State Reconstruction}
\label{AppA}
Here we derive the quantum state reconstruction scheme using the natural gradient \cite{Sorella}, utilising an NNS with two parallel hidden layers as described in Section (II.B). As before we ask the question: How can we optimise the parameters of the NNS, $\alpha = \{ \Omega, \Xi \}$ in order to minimise the ``distance" (maximise the fidelity) between the NNS and the target state? That is
\begin{gather}
\ket{\Psi_{\alpha}(\vec{s})} \xrightarrow[\text{optimise}]{} \ket{\Psi_{\alpha^{\prime}}(\vec{s})} \equiv \ket{\varphi(\vec{s})}, \\
\alpha = \{ \Omega, \Xi \} \xrightarrow[\text{optimise}]{}  \alpha^{\prime} = \{ \Omega^{\prime}, \Xi^{\prime} \},
\end{gather}
where $\alpha$ admits the complete set of parameters for both amplitude and phase learning. Consider an infinitesimally small perturbation $\delta \alpha_k$ performed on the $k^\text{th}$ parameter of the neural network. A linear approximation to the state is given by,
\begin{align}
\ket{\Psi_{\alpha+\delta\alpha_k}}  = \ket{\Psi_{\alpha}} + \sum_k \delta \alpha_k \frac{\partial \ket{\Psi_{\alpha}}}{\partial \alpha_k} =\ket{\Psi_{\alpha}} + \sum_k \delta \alpha_k O_k \ket{\Psi_{\alpha}} 
\end{align}
where the $O_k$ are defined as diagonal matrices whose non-zero elements the partial derivative of the natural logarithm of the NNS vector with respect to the $k^\text{th}$ network parameter \cite{Carleo2}, i.e,
\begin{equation}
O_k = \frac{\partial \ln (\ket{\Psi_{\alpha}})}{\partial \alpha_k}= \frac{1}{\ket{\Psi_{\alpha}}} \frac{\partial \ket{\Psi_{\alpha}}}{\partial \alpha_k}
\end{equation}
In order to reconstruct the target quantum state we wish to force the parameters of the network $\alpha$ to evolve such that the state $\ket{\Psi_{\alpha}}$ is equivalent to that of the target state $\ket{\varphi}$. Therefore, we wish to minimise the ``distance" between the target state and the NNS at every variational perturbation of the parameters. One way of measuring this distance is using the Schatten-2 norm, 
\begin{gather}
\delta =  \left\lVert \ket{\Psi_{\alpha}} + \sum_k \delta \alpha_k O_k \ket{\Psi_{\alpha}} - \ket{\varphi}\right\rVert_2^2
\end{gather}
Through the minimisation of this functional, it is possible to obtain an expression for the perturbation $\delta \alpha_k$ to the $k^\text{th}$ parameter of the network that will optimise the NNS towards the target state. In fact, minimisation of $\delta$ in return of $\delta \alpha_k$  provides a metric,
\begin{equation}
\alpha_k^\prime = \alpha_k +\eta\> \delta\alpha_k 
\end{equation}
where $\eta$ is the learning rate. Using this iterative update rule to the network, with a small enough $\eta$ and enough variational updates, convergence towards the target state should be guaranteed. Defining the operator $\mathcal{F}$,
\begin{gather}
\mathcal{F} = \frac{(\ket{\Psi_{\alpha}}- \ket{\varphi}) \bra{\Psi_{\alpha}}}{\left| \Psi_\alpha \right|^2} 
\end{gather}
one reveals the system of equations for which the minimisation of Eq. (A5) is achieved are
\begin{gather}
\sum_l S_{kl} \delta\alpha_l = f_k \\
S_{kl} = \bra{\Psi_\alpha} O_k^\dag O_l \ket{\Psi_\alpha}  + \bra{\Psi_\alpha}  O_l^\dag O_k \ket{\Psi_\alpha} \\
f_k = \bra{\Psi_\alpha}  \mathcal{F}^\dag O_k \ket{\Psi_\alpha}  + \bra{\Psi_\alpha}  O_k^\dag \mathcal{F} \ket{\Psi_\alpha} 
\end{gather}
Here, $S_{kl}$ denote the elements of a covariance matrix $\mathcal{S}$, and $f_k$ the elements of a generalised force vector $\vec{f}$. By constructing this set of equations, the method of natural gradient descent for the this optimisation scheme can be achieved by defining the metric $\delta\alpha_l$ which is natural to the variational subspace being explored
\begin{equation}
\delta\alpha_l = \sum_k S_{l,k}^{-1} f_k,
\end{equation}
and an update to the variational parameters at every iteration in the optimisation is thence,
\begin{equation}
 \alpha_l^\prime = \alpha_l + \eta \sum_k S_{lk}^{-1} f_k 
\end{equation}
where ${S_{lk}}^{-1}$ denotes the Moore-Penrose pseudo inverse, since this matrix is generally non-invertible. The quantum expectation values throughout $S_{kl}$ and $f_k$ can be efficiently computed as statistical expectation values $\left\langle \cdot\cdot\cdot \right\rangle_\alpha$ according to the probability distribution of the NNS. Rewriting the covariance matrix and generalised forces in this way,
\begin{gather}
S_{kl} \propto \text{Re}{\left\langle O_k^* O_l \right\rangle_\alpha},\>\>
f_{k} \propto \text{Re}{\left\langle O_k^* \mathcal{F} \right\rangle_\alpha},
\end{gather}
they can be incorporated into an iterative numerical method to compute the update rule at every iteration.

\section{Degeneracy of Specific Separabilities for Multipartite States}
\label{AppB}
Consider an $N$-partite state $\ket{\psi}$ whose entanglement properties are described by the set of $K$ disjoint subsets $\mathcal{S} = \{\mathcal{S}_j\}_{j=1,\ldots,K}$ which defines the subsets that contain the indices of potentially entangled qubits (arbitrarily qubits, could be qudits) in the state. A state is deemed $\mathcal{S}$-separable if it is described by this set of exact partitions. This is the most detailed level of entanglement classification we can achieve.\par
Now let $\mathcal{M}$ be the set that defines the size of each of these entangled subsets, i.e  $\mathcal{M}$ for $\mathcal{S}$ is given by  $\{m_j = |\mathcal{S}_j |\}_{j = 1,\ldots,K}$. A state is deemed $\mathcal{M}$-separable if it is described by entangled sub-collections of these dimensions. Given an arbitrary form of $K$-separability we wish to deduce how many forms of $\mathcal{S}$-separability are attributed to it.\par
Whilst $\mathcal{S}$-separability describes a specific separable order of entangled qubits, we can define $\mathcal{M}$-separability as a specific separable order of $m_j$-dimension entangled qubit sets, which is therefore less degenerate than $\mathcal{S}$ (many ordered sets $\mathcal{S}$ may correspond to a single $\mathcal{M}$). By finding the number of ways that a state may be $\mathcal{M}$-separable, we can then use the degeneracy of $\mathcal{M}$ with respect to the creation of $K$ partitions to find total degeneracy.\par
When constructing an entanglement set $(\mathcal{S}, \mathcal{M})$, as each $\mathcal{S}_j$ is filled with indices of entangled qubits, the possible choices of qubits for subsequent subsets diminishes (since they are all disjoint). Hence  for $m_j \in \mathcal{M}$, the number of possible permutations are given by the multinomial coefficient,
\begin{gather}
P = \prod_{i=1}^K \binom{N - \sum_{j=1}^i m_j}{m_i} = \binom{N}{m_1,m_2,\ldots,m_K} = \binom{N}{\mathcal{M}}.
\end{gather}
However, counting in this manner disregards cyclic invariance of separabilities i.e shuffling subsets in $\mathcal{S}$ does not alter the separability of the state. Hence we must further reduce $P$ by removing these duplicates. Such duplicates will only occur whenever the total set contains subsets of equivalent size, $m_i = m_j$ for some $i \neq j \in \{1,\ldots,K\}$. We define the function $g(l) = \sum_{i=1}^k \delta(m_i, l)$ as that which counts the degeneracy of subsets of size $l$, where $\delta$ is the Kronecker delta function.\par
We can then find the number of ways that a $\mathcal{M}$-separable state is $\mathcal{S}$-separable,
\begin{equation}
P_{\mathcal{M}} = \Bigg[\>{\prod_{l=1}^{\tilde{m}} g(l)!}\>\Bigg]^{-1} {\binom{N}{\mathcal{M}}}.
\end{equation}
where $\tilde{m} = \max{(\mathcal{M})}$.
We are now left to find how many ways a $K$-separable state can be constructed using $\mathcal{M}$-separability. This is equivalent to searching for the number of solutions to
$
\sum_{i = 1}^{K} m_i = N,
$
for $m_i \in \mathcal{M}$ and fixed $K$. That is, how many ways can construct a set $\{m_1, \ldots, m_K\}$ such that these elements sum to $N$. The solution to this is given by the partition function of exactly $K$ parts $\mathcal{P}(n,K)$ \cite{Wilf}, which has the generating function,
\begin{equation}
\sum_{n} \mathcal{P}(n,K) x^K = \frac{x^K}{\prod_{i=1}^{K}(1-x^i)}
\end{equation}
and can thus use to determine the degeneracy of $\mathcal{M}$-separability with respect to $K$-separability.
Hence for an $N$-partite state, the number of ways in which we can arrange the $N$-qubits into $K$ entangled collections is given by,
\begin{equation}
G_K =  \sum_{n=1}^{\mathcal{P}(N,K) } \Bigg[\>{\prod_{l=1}^{\tilde{m}_n} g(l)!}\>\Bigg]^{-1} {\binom{N}{\mathcal{M}_n}}.
\end{equation}
Therefore the total number of unique forms of separability (discounting genuine, complete multipartite entanglement) is
$
G = \sum_{K=2}^{N} G_K.
$
This result indeed agrees with the more concise answer to the total number of $\mathcal{S}$-separabilities attributed to an $N$-partite quantum system. This is given by the Bell numbers which can be calculated using Dobi\'{n}ski's formula \cite{Rota},
\begin{equation}
B_N = \frac{1}{e}\sum_{k=0}^\infty \frac{k^N}{k!}.
\end{equation}
It can be seen that the first few Bell numbers do indeed generate the number of $\mathcal{S}$-separabilities for $N$ elements,
\[
B_1 = 1, \> B_2 = 2, \> B_3 = 5, B_4 = 15, B_5 = 52, \ldots.
\]
The Bell numbers count all forms entanglement for an $N$-partite system with respect to $\mathcal{S}$-separability. However they do not detail the degeneracy of the more specific $K$-separability, which is instead given by Eq. (B4).

\end{document}